\documentclass[11pt]{article}

\usepackage{moriond,epsfig}

\def\Journal#1#2#3#4{{#1} {\bf #2}, #3 (#4)}

\def\PLB{{\em Phys. Lett.}  B}

\def\PRD{{\em Phys. Rev.} D}

\def\EPJ{{\em Eur. Phys. J.} C}

\def\JHEP{{\em JHEP} }

\def\be{\begin{equation}}

\def\ee{\end{equation}}

\def\bea{\begin{eqnarray}}

\def\eea{\end{eqnarray}}

\begin{document}

\vspace*{4cm}

\title{MEASUREMENTS OF PROTON STRUCTURE FUNCTIONS, $\alpha_s$ AND PARTON DISTRUBUTION FUNCTIONS AT HERA}

\author{ N. WERNER \\ (on behalf of the H1 and ZEUS collaborations) }

\address{Physik-Institut der Universit\"at Z\"urich, Winterthurerstrasse 190, 8057 Z\"urich, Switzerland }

\maketitle 
\abstracts{
The measurement and QCD analysis of neutral and charged current cross sections at HERA are presented.
For the kinematic range of four-momentum transfer $Q^2$ between $\approx$ 1 $GeV^2$ and 30000 $GeV^2$ and Bjorken x
between 0.0013 and 0.65 the data are compared to NLO QCD predictions. In dedicated NLO QCD analyses the extraction
of the parton distribution functions of the proton is performed.
Results for the measurement of the strong coupling constant $\alpha_s$ are presented.}

\section{Introduction}\label{subsec:intro}

Deep inelastic scattering (DIS) experiments have played a major role in the 
understanding of the partonic structure of matter. In the past decade proton structure has been 
studied in $ep$ collisions by the two HERA experiments H1 and ZEUS.\\
Both experiments have been delivering increasingly precise structure 
function measurements leading to improved knowledge of the parton distribution functions (PDFs) 
inside the proton. Measurements 
of the proton structure are a good testing ground for QCD predictions. The precise 
knowledge of the PDFs is also important for reliable predictions of cross sections of the standard model 
and for new physics searches at future hadron colliders such as 
the LHC.\\

\section{The Neutral and Charged Current Cross Sections}\label{subsec:x-sections}

In deep inelastic $ep$ scattering the structure of the proton can be probed by $\gamma$ and $Z^0$ 
or by $W^{\pm}$ exchange. The first are referred to as neutral current (NC) processes 
($ep \rightarrow eX$) whereas the
latter are charged current (CC) interactions ($ep \rightarrow \stackrel{(-)}{\nu} X$).\\
 The NC cross 
section can be expressed in terms of three structure 
functions ${F_2}$, $x{F_3}$ and ${F_L}$:\\
\begin{equation}\begin{array}{lcr}
\frac{d^{2}\sigma_{NC}^{e^{\pm}p}}{dxdQ^{2}}=\frac{2 \pi \alpha^{2}}{xQ^{4}} \cdot 
\Phi_{NC}^{e^{\pm}p},&
 with&
\Phi_{NC}^{e^{\pm}p}=Y_{+} {F_2} \mp
{Y_{-} x {F_{3}}}- y^2 {F_{L}}.\\
\end{array} 
\label{eq:NCxsection1}
\end{equation}
$Y_{\pm}=1 \pm (1-y)^2$ contains the helicity dependences of the electro-weak interactions where 
$y=Q^2/xs$. The negative four-momentum transfer 
squared $Q^2$, 
carried by the exchanged gauge boson, represents its resolution power with respect to 
the proton. The fraction of the proton momentum carried by the
struck parton is given by $x$-Bjorken and $s$ is the center-of-mass energy squared.\\
The structure functions are sensitive to the parton distributions. In the quark parton model, $F_2$ is 
related to the sum of quark and anti-quark momentum distributions, $xq(x,Q^2)$ and $x \bar{q}(x,Q^2)$, 
whereas $xF_3$ is related to their difference. The longitudinal structure function $F_L$ is vanishing in this model.
In the bulk of the
kinematic region at HERA the dominating contribution to the cross section is due to $F_2$
related to the pure photon exchange. Only at large values of $Q^2$ the contribution from 
$Z^0$ exchange becomes important. The longitudinal structure function $F_L$
is significant at large $y$ only.

The inclusive CC cross section for $ep$ interactions can be expressed in terms of 
three structure functions $W_{2}$, $x{W_{3}}$ and $W_{L}$ :\\
\begin{equation} \begin{array}{lcr}
\frac{d^2\sigma^{e^{\pm}}_{CC}}{dx dQ^2}=\frac{G^{2}_{F}}{2\pi x}\bigg(\frac{M_{W}^{2}}
{Q^{2}+M_{W}^{2}}\bigg)^2 \phi_{CC}^{e^{\pm}},&
with &
\phi_{CC}^{e^{\pm}}=\frac{1}{2}\bigg(Y_{+}W_{2}^{\pm}\mp Y_{-}xW_{3}^{\pm}-
y^2 W_{L}^{\pm}\bigg),\\
\end{array}
\end{equation}\label{eq:CCxsection1}\\
where $G_F$ is the Fermi coupling constant.
In leading order (LO) these structure functions are, depending on the charge of the beam lepton,
sensitive to certain quark compositions.\\

\begin{figure}[h]
\begin{center}
\setlength{\unitlength}{1cm}
\begin{picture}(6.5,6)
\put(0,0.){\epsfig{file=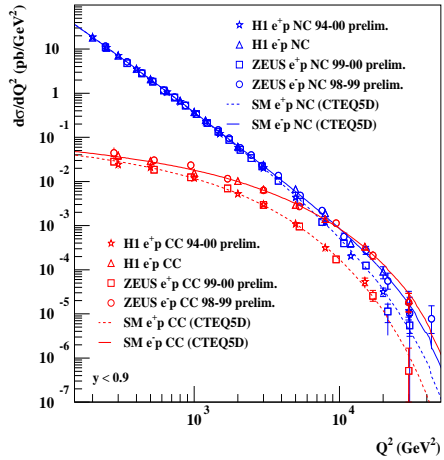,width=6cm,height=6cm,
bbllx=13pt,bblly=43pt,bburx=521pt,bbury=521pt}}
\end{picture}
\end{center}
\caption[]
{\label{plot:x-sections} \it NC and CC cross sections $d\sigma/dQ^2$ measured by H1 and ZEUS
in $e^+p$ and $e^-p$ interactions are compared with the SM predictions based on PDFs from CTEQ5D \cite{cteq5}.}
\end{figure}

The results for the measurements of NC and CC single differential cross sections $\frac {d\sigma}{dQ^2}$ by H1 and ZEUS are
summarised in fig. \ref{plot:x-sections}. 

In NC $\frac{d\sigma}{dQ^2} \sim \frac{1}{Q^4}$ due to the 
dominance of photon exchange. At $Q^2 \approx 100$ GeV$^2$ the cross section is about a factor of 1000 larger than the CC cross section. 
However, at \mbox{$Q^2 \stackrel{>}{\sim} M_{z}^2, M_W^{2}$} both
cross sections are of the same order of magnitude, illustrating electro-weak 
unification in DIS. In $e^-p$ interactions both cross sections are larger than in the case of $e^+p$ 
scattering. In NC the positive (negative) $\gamma Z$ interference term in the $e^-p$ ($e^+p$) interactions 
is responsible for this effect. It can be employed to measure $xF_3$ \cite{xf3}. The difference in the CC cross section is 
caused by the different type of valence quark taking part in the interaction.\\
The measurements of NC and CC $e^{\pm}p$ scattering cross sections provide complementary sensitivity to different
quark distributions and the gluon distribution $xg(x,Q^2)$. The accuracy and kinematic coverage of the cross section data
enables dedicated QCD analyses.

\section{NLO QCD Analysis of HERA data}

\begin{figure}[h]
\begin{center}
\vspace*{13pt}
\setlength{\unitlength}{1cm}
\begin{picture}(16,9)
\put(-0.5,0.){\epsfig{file=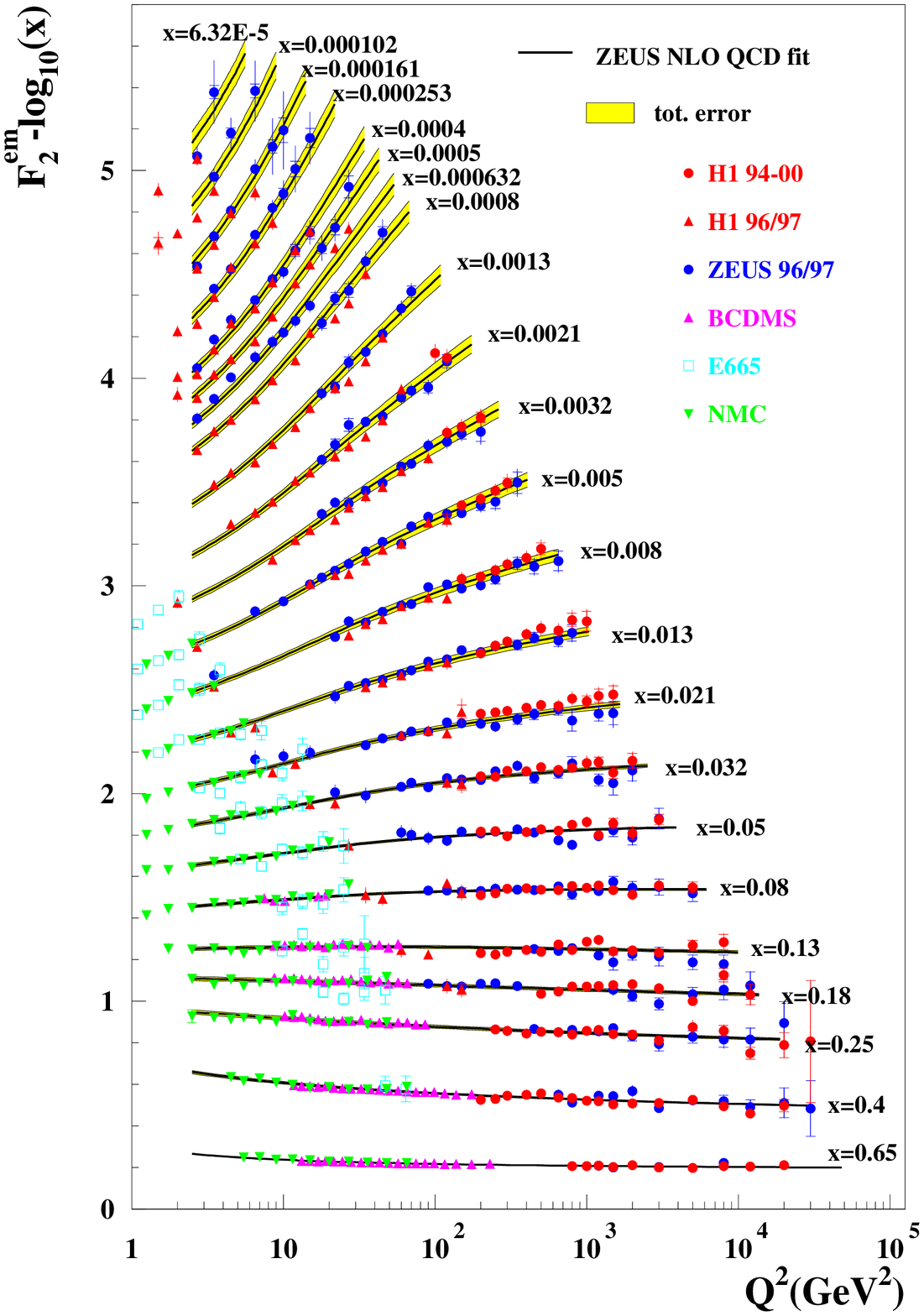,width=7cm,height=9cm,
bbllx=12pt,bblly=55pt,bburx=534pt,bbury=775pt}}
\put(8,0.4){\epsfig{file=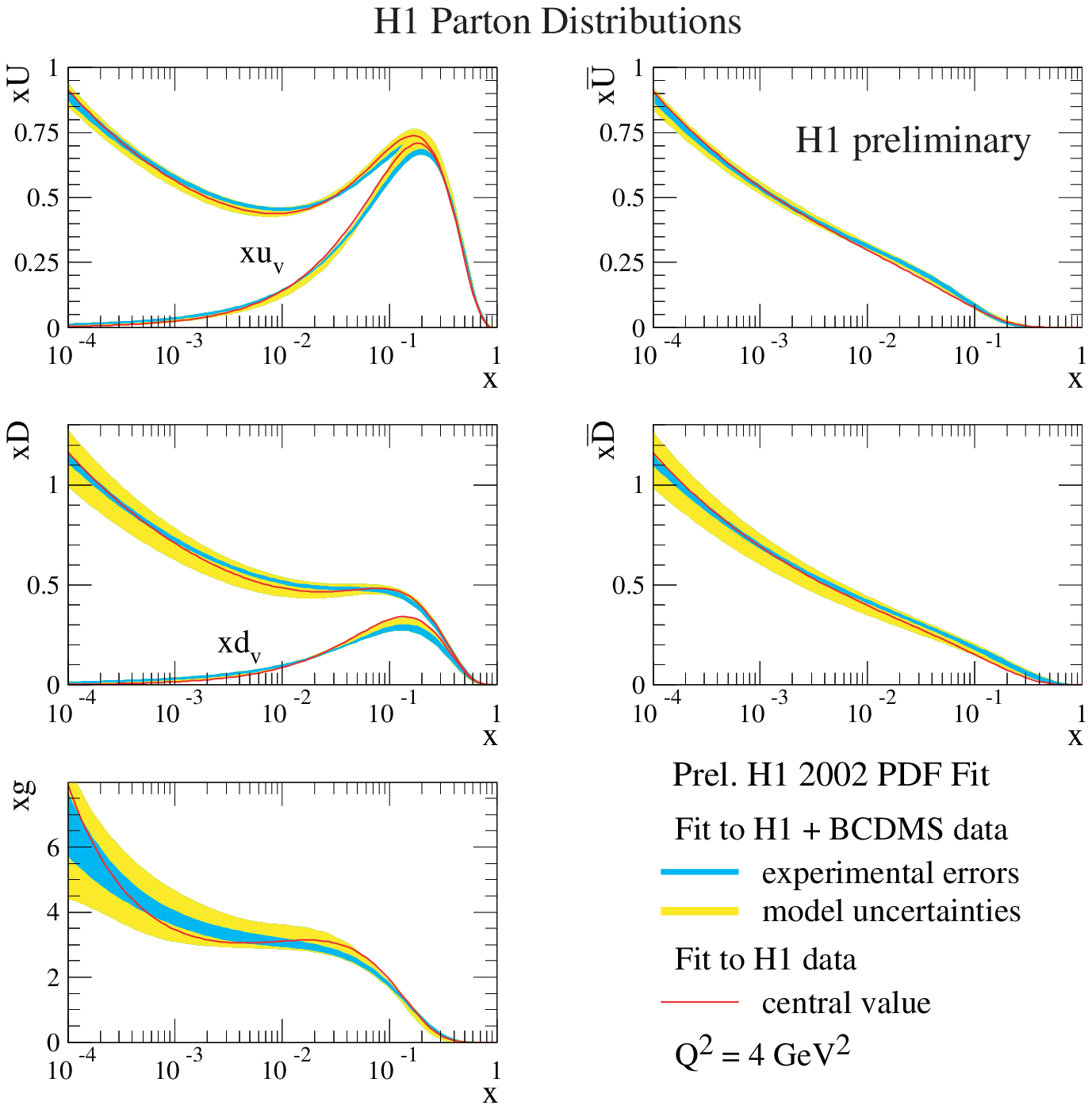,width=8cm,height=9cm,
bbllx=95pt,bblly=219pt,bburx=500pt,bbury=619pt}}
\end{picture}
\caption[]
{\label{plot:F2-fit} \it Left: $F_{2}^{em}$, i.e. $F_2$ from pure $\gamma$ exchange, from 
HERA and fixed target experiments compared with ZEUS NLO QCD fit. Right: PDFs $xU=x(u+c)$, $x\bar{U}=x(\bar{u}+\bar{c})$, $xD=x(d+s)$, 
$x\bar{D}=x(\bar{d}+\bar{s})$ and xg as determined from the H1 fits at the starting scale $Q_0^2 = 4$ GeV$^2$.}
\end{center}
\end{figure}

In most of the kinematic range the NC cross section is dominated by the structure function $F_2$.
Using Eq. \ref{eq:NCxsection1} $F_2$ can be extracted from the measurement of the
double differential NC cross section. 
Fig. \ref{plot:F2-fit} (left) shows the H1 and ZEUS $F_2$ data 
together with data from fixed target experiments. The data are well described by the DGLAP evolution over four orders of 
magnitude in $x$ and $Q^2$. 
The steep rise of $F_2$ at small $x$ is driven by the gluon.\\
In the NLO QCD fit the PDFs are parameterised at a fixed starting scale
$Q^2=Q_0^2$ and determined by the fitting procedure. The resulting PDFs at the starting scale 
obtained by H1 are shown in fig. \ref{plot:F2-fit} (right). The present accuracy of the NC and CC 
HERA data allows to perform a fit to HERA data alone.

\begin{figure}[h!!]
\begin{center}
\setlength{\unitlength}{1cm}
\begin{picture}(12,7)
\put(-1,0.){\epsfig{file=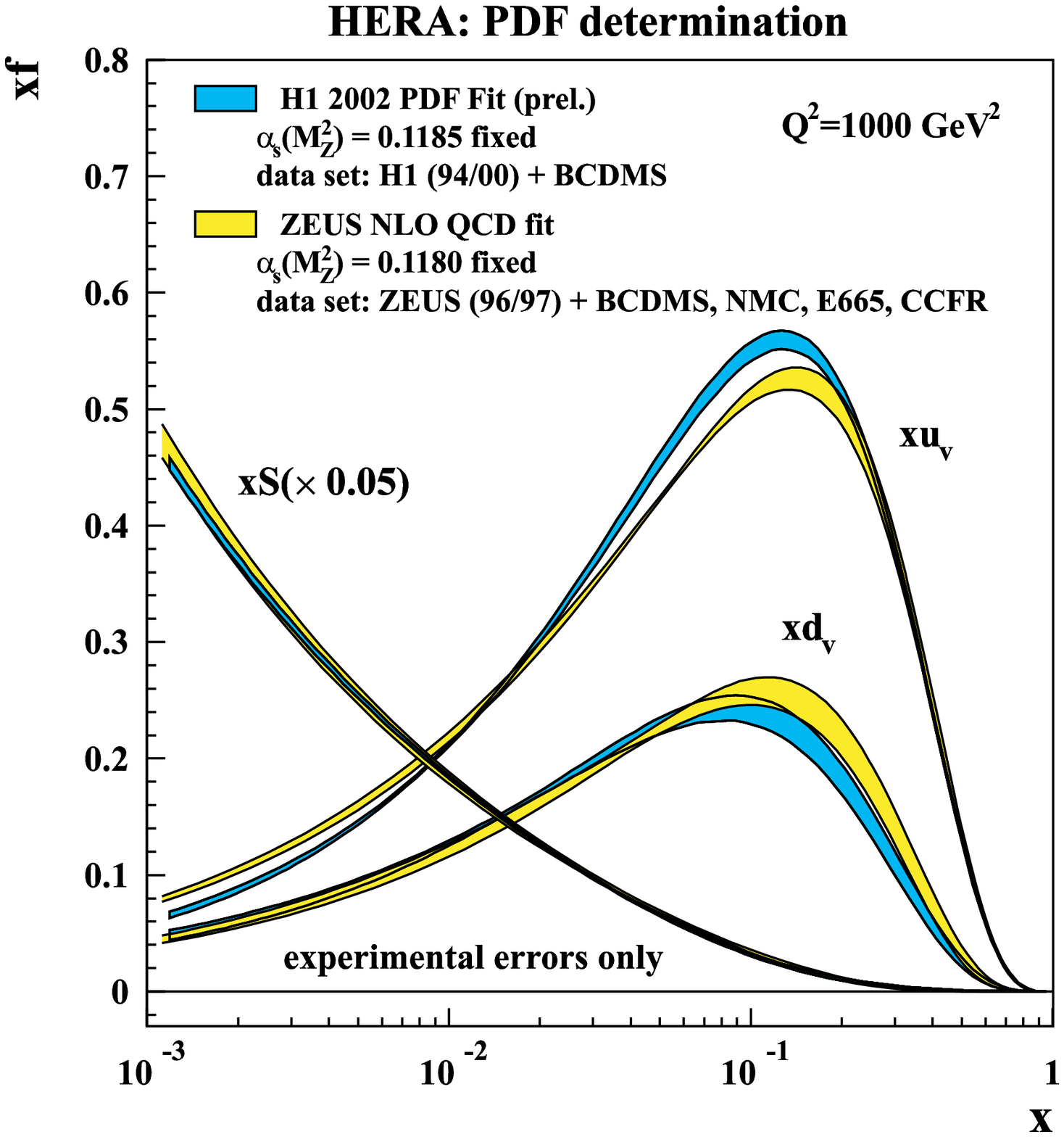,width=6.5cm,height=6.5cm,bbllx=37pt,bblly=164pt,bburx=547pt,bbury=681pt}}
\put(6,0.){\epsfig{file=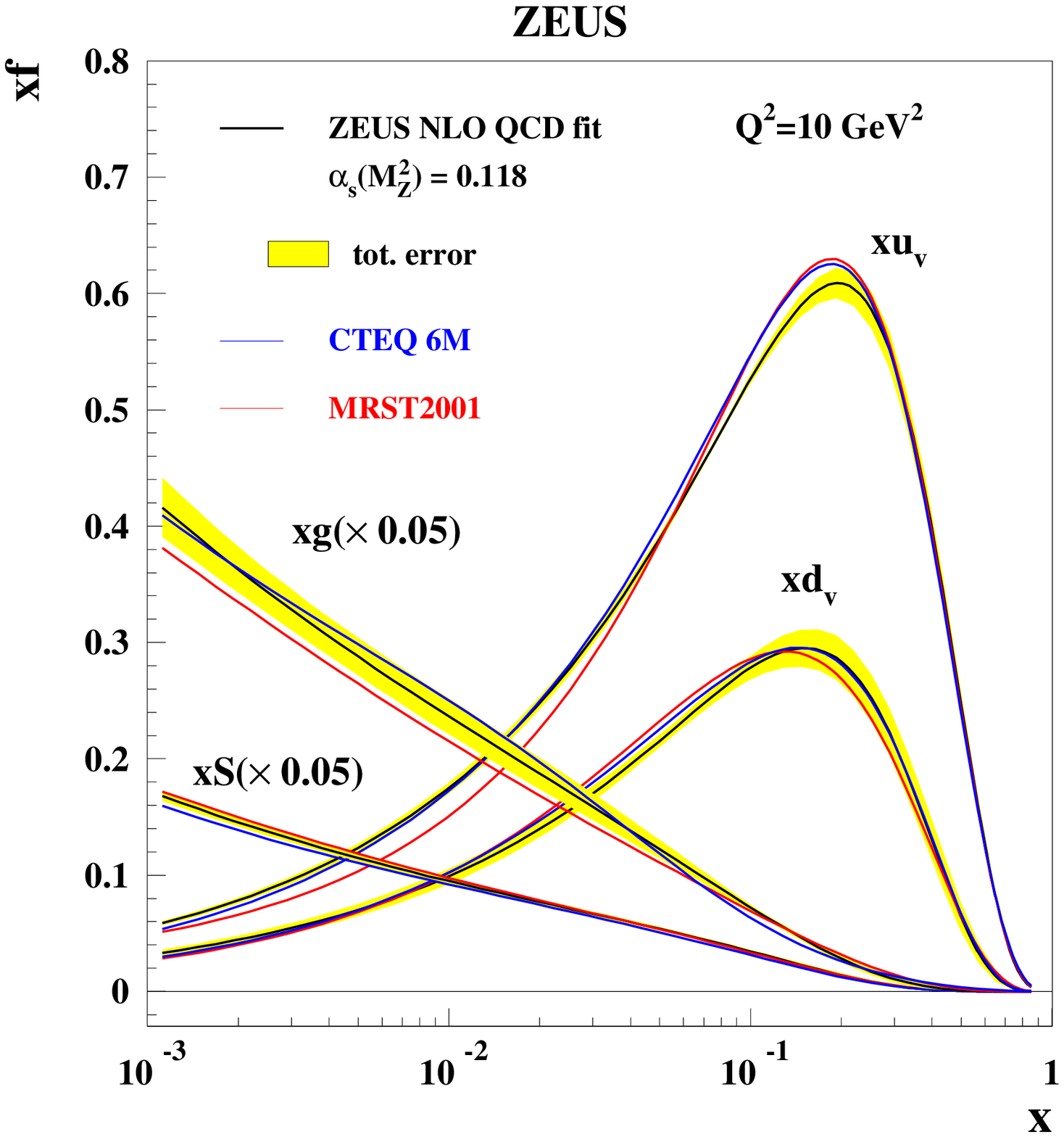,width=6.5cm,height=6.5cm,bbllx=37pt,bblly=164pt,bburx=547pt,bbury=681pt}}
\end{picture}
\caption[]
{\label{plot:pdfs} \it Left: Comparison of the PDFs extracted by H1 \cite{h1} and ZEUS \cite{zeus}. 
Right: Comparison of the PDFs from the ZEUS fit to global fits by MRST2001 \cite{mrst} and CTEQ6M \cite{cteq}}
\end{center}
\end{figure}

The evolution of the PDFs in $Q^2$ at fixed $x$ is predicted by the NLO DGLAP 
\mbox{equations \cite{dglap}.}
The PDFs at $Q^2 = 1000$ GeV$^2$ as determined from the NLO QCD fits are shown for H1 \cite{h1} and \mbox{ZEUS \cite{zeus}} 
in fig. \ref{plot:pdfs} (left). 
At $Q^2 = 10$ GeV$^2$ the ZEUS results are also compared to global fits (right).
The results of H1 and ZEUS are consistent with the
results of the global analyses MRST2001 \cite{mrst} and CTEQ6M \cite{cteq}.  \\


In dedicated QCD fits $\alpha_s(M_z)$ has been measured by both experiments. The results obtained by H1 and ZEUS
are $\alpha_s(M_z)=0.1150\pm
0.0017(exp)^{+0.0009}_{-0.0007}(model)$ \cite{h1-alphas} and $\alpha_s(M_Z)=0.1166\pm0.0008(uncorr.)\pm 0.0032(corr.) \pm 0.0036(norm.) 
\pm 0.0018(model)$ \cite{zeus} respectively. Both results are consistent and competitive with the world average.
However, theoretical uncertainties due to higher order effects are estimated to be $\pm 0.005$. For a reduction of these errors
NNLO calculations are indispensable.

\section{Conclusion}
H1 and ZEUS have performed NLO QCD analyses of their data. The
data are well described by the predictions. Extractions of the parton densities and a measurement of the strong coupling constant 
$\alpha_s$ were performed.\\
In order to improve the theoretical uncertainties of the measurements NNLO calculations are
needed.\\
With the HERA luminosity upgrade the amount of data will increase substantially and polarisation of the lepton beams will be achieved.
More precise and new measurements will thus be possible in the future and help to improve the understanding of the 
structure of the proton.

\end{document}